\documentclass[a4paper,11pt]
{article}

\usepackage{jheppub}

\usepackage{slashed}
\usepackage{wrapfig}
\usepackage{bm}
\usepackage{latexsym,amssymb,amsmath,float,url,mathrsfs}
\usepackage{latexsym}
\usepackage{graphicx}
\usepackage{epstopdf}
\usepackage{amsfonts}
\usepackage{amsmath}
\usepackage{amssymb}
\usepackage{comment}
\usepackage{subfigure}
\usepackage{natbib}
\usepackage{hyperref}
\usepackage{pifont}
\usepackage{blindtext}
\usepackage{adjustbox}
\usepackage{multirow}
\usepackage{tabularx}
\usepackage[table]{xcolor}
\usepackage{orcidlink}
\usepackage{tikz}
\usetikzlibrary{tikzmark}
\usepackage{amssymb}
\usepackage{bbold}
\usepackage{orcidlink}

\usepackage{blkarray}
\usepackage{graphicx}
\usepackage{amsfonts}
\usepackage{soul}
\usepackage{amssymb}
\usepackage{amsmath}
\usepackage{cancel}
\usepackage{tcolorbox}

\def\BH{\rm B\text{-}H}

\def\MSbar{\relax\ifmmode\overline{\rm MS}\else{$\overline{\rm MS}${ }}\fi}

\def\BH{{\rm{B\!-\!H}}}
\def\bh{{\rm bh}}
\usepackage{comment}

\usepackage{graphics,appendix,afterpage,makecell} 

\definecolor{oucrimsonred}{rgb}{0.6, 0.0, 0.0}
\definecolor{persianblue}{rgb}{0.11, 0.22, 0.73}
\definecolor{forestgreen}{rgb}{0.13,0.35,0.13}
\definecolor{lightgray}{rgb}{0.83, 0.83, 0.83}
 \hypersetup{colorlinks, citecolor=oucrimsonred, linkcolor=black, urlcolor=oucrimsonred}
\definecolor{cornellred}{rgb}{0.7, 0.11, 0.11}
\definecolor{navyblue}{rgb}{0.0, 0.0, 0.5}
\definecolor{amethyst}{rgb}{0.6, 0.4, 0.8}
\definecolor{yellow}{rgb}{1.0, 1.0, 0.0}
\definecolor{firebrick}{rgb}{0.7, 0.13, 0.13}
\definecolor{tangerineyellow}{rgb}{1.0, 0.8, 0.0}
\definecolor{deepfuchsia}{rgb}{0.76, 0.33, 0.76}
\definecolor{amber}{rgb}{1.0, 0.75, 0.0}
\definecolor{VioletRed4}{rgb}{0.55, 0.13, .32}
\definecolor{indiagreen}{rgb}{0.07, 0.53, 0.03}
\definecolor{VioletRed4}{rgb}{0.55, 0.13, .32}
\newcommand{\be}{\begin{equation}}
\newcommand{\ee}{\end{equation}}
\newcommand{\bea}{\begin{equation} \begin{aligned}}
\newcommand{\eea}{\end{aligned} \end{equation}}

\definecolor{oucrimsonred}{rgb}{0.6, 0.0, 0.0}
\newcommand\vertarrowbox[3][6ex]{%
  \begin{array}[t]{@{}c@{}} #2 \\
  \left\uparrow\vcenter{\hrule height #1}\right.\kern-\nulldelimiterspace\\
  \makebox[0pt]{\scriptsize#3}
  \end{array}%
}

\definecolor{verdechiaro}{rgb}{0.6,1,0.6}
\definecolor{giallochiaro}{rgb}{1,1,0.6}
\definecolor{bluscuro}{rgb}{0.15, 0.2, 0.9}
\definecolor{verdes}{rgb}{0.1, 0.5, 0.1}%
\definecolor{tangerineyellow}{rgb}{1.0, 0.8, 0.0}

\definecolor{americanrose}{rgb}{1.0, 0.01, 0.24}
\definecolor{cobalt}{rgb}{0.0, 0.28, 0.67}
\definecolor{brandeisblue}{rgb}{0.0, 0.44, 1.0}
\definecolor{mycolor}{rgb}{0.0, 0.0, 0.5}
\definecolor{oxfordblue}{rgb}{0.0, 0.13, 0.28}
\definecolor{azure}{rgb}{0.0, 0.5, 1.0}
\definecolor{turquoiseblue}{rgb}{0.0, 1.0, 0.94}
\newtcolorbox{mynewbox}[1]{colback=white!5!white,colframe=azure!75!black,fonttitle=\bfseries,title=#1}
\newtcolorbox{mybox}{colback=mycolor!5!white,colframe=azure!75!black}
\newtcolorbox{mynamedbox}[1]{colback=mycolor!5!white,colframe=azure!75!black,title=#1}
\definecolor{venetianred}{rgb}{0.78, 0.03, 0.08}
\newtcolorbox{mynamedbox1}[1]{colback=venetianred!5!white,colframe=venetianred!80!black,title=#1}
\newtcolorbox{mynamedbox2}[1]{colback=azure!5!white,colframe=azure!80!black,title=#1}

\definecolor{verdes}{rgb}{0.1, 0.5, 0.1}%
\definecolor{cornellred}{rgb}{0.7, 0.11, 0.11}

\definecolor{VioletRed4}{rgb}{0.55, 0.13, .32}

\hypersetup{
     colorlinks   = true,
     citecolor    = violet,
     urlcolor     = violet,
     linkcolor    = violet}

\definecolor{rossocorsa}{rgb}{0.83, 0.0, 0.0}

\usepackage[normalem]{ulem}

\vspace*{1.5cm}

\title{
Information--Theoretic Black Hole Entropy II: Infrared Gravity  and 
Charged/Rotating Extensions}

\author[a]{Alex Kehagias\orcidlink{0000-0001-6080-6215}}
\affiliation[a]{Physics Division, National Technical University of Athens, Athens, 15780, Greece}

\abstract{
We investigate gravitational and Kerr--Newman extensions of an
information-theoretic black-hole entropy that satisfies the Nernst
formulation of the third law of thermodynamics. The entropy is identified with the
Kullback--Leibler divergence between a mass-dependent Bernoulli ensemble
and an unbiased reference ensemble, and therefore measures relative
information rather than the logarithm of the number of black-hole
microstates. We show perturbatively that its temperature and entropy can be
reproduced by an infrared deformation of General Relativity. To linear
order in the deformation couplings, the surface-gravity temperature and
the Wald entropy agree with the information-theoretic results. In an
explicit realization, the matching generates a positive effective
cosmological term whose smallness is related to the large microscopic
parameter $N$. We also propose an extension to Kerr--Newman black holes
based on the irreducible mass. This construction preserves the connection
with the horizon area and the semiclassical limit, while distinguishing
geometrical extremality from the universal statistical-freezing endpoint,
which is independent of angular momentum and charge.

}

\emailAdd{kehagias@mail.ntua.gr}

\makeatletter
\gdef\@fpheader{}
\makeatother

\begin{document}
\maketitle

\section{Introduction}

Black hole thermodynamics provides one of the clearest indications that gravity, quantum theory, and statistical mechanics belong to a unified framework. Once the horizon area is associated with entropy and surface gravity with temperature \cite{Bekenstein:1973ur,Bardeen:1973gs,BekensteinS,Hawking:1975vcx}, the laws of black hole mechanics acquire a clear thermodynamic interpretation. Although a wide range of semiclassical results supports this correspondence, its conceptual foundations remain incomplete.
The central unresolved issue is the microscopic meaning of black hole entropy. While the entropy of ordinary thermodynamic systems is determined by a probability distribution over microscopic configurations, gravitational entropy is expressed instead through the geometric area of the horizon. Consequently, it remains unclear what the underlying degrees of freedom are, where they are localized, or how precisely the area law measures missing information.

Several frameworks have provided important partial answers to these questions.
For instance, string/M-theory reproduces the area law for specific
supersymmetric black holes through microscopic state counting
\cite{Strominger1996,Maldacena:1997de}.
The quantity that remains theoretically manageable in many related constructions is a protected supersymmetric index. While these results provide strong evidence for a microscopic origin of black hole entropy within specific sectors, they do not by themselves establish a universal state-counting interpretation for generic black holes. 
Alternative approaches have explored such issues, including including string- and field-theoretic approaches to horizon entropy \cite{Strominger1996,Strominger1997,Dvali2024,Susskind1994,Susskind1994-2} as well as the quantum $N$-portrait, which describes black holes as self-sustained Bose-Einstein condensates \cite{Dvali2011,Dvali2015}. Area-law scaling of a genuine microstate entropy has also been shown to arise in a nongravitational system at quantum criticality \cite{Dvali:2017nis}.

In this work and in its companion paper \cite{kehagias-1}, we focus our attention specifically on the third law \cite{wilks1961}. The Bekenstein-Hawking area law \cite{Bekenstein:1973ur} endows black holes with an entropy that aligns beautifully with the first and second laws of thermodynamics \cite{Bardeen:1973gs,BekensteinS,Hawking:1975vcx}. However, it violates the third law because as the Hawking temperature $T \to 0$, which formally occurs as $M \to \infty$ for a Schwarzschild black hole, the entropy diverges instead of approaching a universal constant as required by the Nernst formulation of the thord law \cite{wilks1961}. This distortions and apparent conflicts have been thoroughly discussed in the literature \cite{Bardeen:1973gs,Wald1,Wald2,Israel,Reall1,Reall2}, with various perspectives on whether the third law should hold for black holes within General Relativity. Recent constructions provide counterexamples to the unattainability
formulation of the third law, including  the gravitational collapse in the
Einstein--Maxwell charged-scalar system and the finite-time extremal black hole
formation in five-dimensional vacuum gravity
\cite{KU,Crump:2026kgu}.

In our companion paper \cite{kehagias-1}, we argued that the third law should also be satisfied. This requirement, combined with the established high-temperature and low-mass behavior of the Bekenstein-Hawking entropy, led us to a modified entropy function $S_{\rm bh}(M)$, which takes a maximal constant value  at $T=0$. This derivation proceeds by analogy with the blackbody case, where Planck's formula corrects the naive high-energy entropy $S \sim \ln E$ once a minimum energy ground state is introduced. For black holes, the corresponding modification introduces a maximum mass  $M_0$ that acts as a universal scale.

The present paper extends the analysis of \cite{kehagias-1} in two distinct directions by addressing two subsequent questions that arise naturally from that construction.
\vspace{2mm}

\noindent\textbf{Question 1.} Since  the Bekenstein-Hawking entropy in General Relativity does not satisfy the Nernst third law, what modification of the gravity theory is required to reproduce $S_{\rm bh}(M)$?
\vspace{2mm}

\noindent
and
\vspace{2mm}

\noindent\textbf{Question 2.} How does the entropy formula generalize to rotating and/or charged (Kerr-Newman) black holes?\vskip2mm

We show that $S_{\rm bh}(M)$ arises from a class of infrared-deformed gravitational theories and that the natural extension to the Kerr-Newman sector is obtained by replacing $M$ with the irreducible mass $M_{\rm irr}$, which is tied directly to the horizon area and remains invariant under reversible extraction processes. The remainder of this paper is organized as follows. Section~\ref{sec:summary} provides a concise summary of the main results from the companion work, which include the derivation of $S_{\rm bh}(M)$, its microscopic interpretation as an entropy deficit, and its identification as a Kullback-Leibler divergence \cite{KullbackLeibler1951,CoverThomas2006}. Section~\ref{sec:further} then details our new contributions by developing the modified gravitational theory and Section~\ref{sec:extension} describes the explicit Kerr-Newman extension. Section~\ref{sec:CC} presents the connection of the universal mass scale $M_0$ with the cosmological constant, and finally, Section~\ref{sec:conclusions2} contains our concluding remarks.


\section{Summary of the companion paper}
\label{sec:summary}

We summarize here the key results derived in \cite{kehagias-1}, covering the
derivation of the modified entropy formula, its microcanonical interpretation,
and its identification as a Kullback-Leibler divergence.

\subsection{Black Hole Entropy consistent with the third law}

The standard Bekenstein-Hawking entropy $S_{\rm B\text{-}H}$ for a Schwarzschild black hole of
mass $M$ is (in $G_N = 1$ units) 
\begin{eqnarray}
    S_{\rm B\text{-}H} = 4\pi M^2, \label{BH}
\end{eqnarray}
with
corresponding Hawking  temperature 
\begin{eqnarray}
    T_H = \frac{1}{8\pi M}.
    \label{TH}
\end{eqnarray}
  As $T_H$ approaches zero (i.e.\
$M \to \infty$), the entropy diverges ($S_\BH\to \infty)$, in violation of the Nernst's 
third law of thermodynamics \cite{wilks1961}, which requires the entropy to approach a universal constant at vanishing temperature. 
Likewise, the heat capacity
\begin{eqnarray}
    C_V = -\frac{1}{8\pi T_H^2},
\end{eqnarray}
 diverges instead of vanishing as $T \to 0$.

To see how the third law might be restored, we can compare with the case  of blackbody radiation. 
For a blackbody system, 
the naive high-energy entropy $S \sim \ln E$ violates the Nernst theorem, but the full Planck formula restores it by introducing a minimum energy $E_0$ representing the zero-point energy.  For black holes, the analogous step is to introduce a
\emph{maximum mass scale} $M_0$, such that $M\leq M_0$, at which 
\begin{eqnarray}
    T(M_0)=0, \qquad \mbox{and}\qquad S_{\rm bh}(M_0)=S_0,
\end{eqnarray}
where  $S_0$ is a universal constant. 
Imposing then that the second derivative $d^2 S_{\rm bh}/dM^2$ has only a simple
pole at $M = M_0$ (higher-order poles produce divergences in $S_{\rm bh}$), demanding
the correct high temperature (Bekenstein-Hawking) limit, and fixing integration
constants, one finds \cite{kehagias-1}
\begin{equation}
  S_{\rm bh}(M) = 4\pi M_0^2
  \bigg[\left(1-\frac{M}{M_0}\right)\ln\!\left(1-\frac{M}{M_0}\right)
       +\left(1+\frac{M}{M_0}\right)\ln\!\left(1+\frac{M}{M_0}\right)\bigg].
\label{sm3}
\end{equation}
The corresponding temperature and heat capacity turn out then to be
\begin{equation}
  \frac{1}{T} = 4\pi M_0 \ln\!\left(\frac{M_0+M}{M_0-M}\right),
\label{temp}
\end{equation}
\begin{equation}
  C_V = -\frac{1}{8\pi T^2}\,\mathrm{sech}^2\!\left(\frac{1}{8\pi M_0 T}\right).
\label{BHc}
\end{equation}
The entropy \eqref{sm3} can be expanded in powers of $M/M_0 $ as 
\begin{eqnarray}
    S_{\rm bh}=4\pi M^2+\frac{2\pi}{3}\frac{M^4}{M_0^2}+\cdots ,
\end{eqnarray}
and similarly,  the corresponding temperature and heat capacity 
\begin{eqnarray}
    T= \frac{1}{8\pi M}-\frac{M}{24\pi M_0^2}+\cdots \, ,\qquad \mbox{and}\qquad 
    C_V= -\frac{1}{8\pi T^2}+\frac{1}{512 \pi^3 M_0^2 T^4}+\cdots\, .
\end{eqnarray}
Therefore, the Bekenstein-Hawking entropy is recoved in the $M_0\to \infty$ limit. In addition the entropy $S_{\rm bh}$
vanishes at $M = 0$ and reaches a maximum  at
$M = M_0$, where it takes the value
\begin{eqnarray}
    S_{\rm max} = 8\pi M_0^2 \ln 2.
\end{eqnarray}
Notice that 
\begin{eqnarray}
   T \xrightarrow[M \to M_0]{} 0,\qquad  S_{\rm bh} \xrightarrow[T \to 0]{} S_0=S_{\rm max}
\end{eqnarray}
in accordance with the third law.  

It is convenient to introduce, after restoring the reduced Planck mass $M_P^2=1/8\pi G_N$, the parameter  
\begin{eqnarray}
   N = \frac{M_0^2}{M_P^2 }.
   \label{eq:N}
\end{eqnarray}
Then, the
entropy in Eq. (\ref{sm3}) can be written as
\begin{equation}
  S_{\rm bh} = \frac{N}{2}
  \bigg[\left(1-\frac{M}{M_0}\right)\ln\!\left(1-\frac{M}{M_0}\right)
       +\left(1+\frac{M}{M_0}\right)\ln\!\left(1+\frac{M}{M_0}\right)\bigg],
\label{sm5}
\end{equation}
which admits an  $1/N$ expansion as
\begin{equation}
  S_{\rm bh} = \frac{M^2}{2M_P^2}
  \left(1 + \frac{1}{N}\frac{M^2}{6M_P^2} + \frac{1}{N^2}\frac{M^4}{15M_P^4}
  + \cdots\right).
\label{SN}
\end{equation}
Clearly, the leading term is the Bekenstein-Hawking entropy.

\subsection{Typical-set interpretation of the entropy}
As  we have shown in \cite{kehagias-1}, the entropy \eqref{sm5} admits a transparent statistical-mechanical
interpretation.  Introducing the parameter
\begin{eqnarray}
    p = \frac{1}{2}\left(1 +\frac{ M}{M_0}\right),
\end{eqnarray}
one can rewrite $S_{\rm bh}$ as
\begin{equation}
  S_{\rm bh} = N \ln 2 - N H(p),
\end{equation}
where
\begin{eqnarray}
   H(p) = -p \ln p - (1-p)\ln(1-p) 
\end{eqnarray}
is the Shannon entropy of a
Bernoulli distribution with bias $p$.  This shows that $S_{\rm bh}$ is an
\emph{entropy deficit}, and  measures the gap between the maximal entropy
$S_{\rm max}=\ln 2^N$ of $N$ unbiased bits and the Shannon entropy $N H(p)$ accessible to
an exterior observer who can only probe the biased statistics encoded in
$p(M)$.

Equivalently, this can be expressed by defining $W_{\rm tot} = 2^N$ as the total number of microstates of $N$ unbiased bits. The number of typical ``visible'' configurations consistent with a bias $p$ is then given by $W_{\rm vis} = 2^{N H_2(p)}$, where $H_2(p)$ is the Shannon entropy in bits. With these definitions, one finds
\begin{equation}
S_{\rm bh} = N\ln 2\Big[1-H_2(p)\Big]=\ln\left(\frac{W_{\rm tot}}{W_{\rm vis}}\right).
\end{equation}
Thus, $S_{\rm bh}$ remains the thermodynamic black-hole entropy, but in the
present Bernoulli representation it takes the form of an entropy deficit
relative to the maximally mixed reference ensemble.

\subsection{Black hole entropy as Kullback-Leibler divergence}

The entropy deficit interpretation becomes most transparent when $S_{\rm bh}$
is written in terms of the Kullback-Leibler (KL) divergence.  For the
single-bit biased distribution 
\begin{eqnarray}
    p_\pm = \frac{1}{2}\left(1 \pm \frac{M}{M_0}\right),
    \label{ppm}
\end{eqnarray}
 the  KL divergence relative
to the uniform reference distribution $q = \tfrac{1}{2}$ is

\begin{equation}
  D_{\rm KL}\!\left(p\Big\Vert\frac{1}{2}\right)
  = \sum_{i=\pm} p_i \ln \frac{p_i}{1/2}
  = p_+ \ln (2p_+) + p_- \ln (2p_-),
\end{equation}
which, using the bias $p_\pm$ of Eq. \eqref{ppm}, is written as 
\begin{equation}
  D_{\rm KL}\!\left(p\Big\Vert\frac{1}{2}\right)
  = \frac{1}{2}\bigg[
    \left(1-\frac{M}{M_0}\right)\ln\!\left(1-\frac{M}{M_0}\right)
  + \left(1+\frac{M}{M_0}\right)\ln\!\left(1+\frac{M}{M_0}\right)
  \bigg].
\end{equation}
For an ensemble of $N$ independent such bits, the total KL divergence is
$N\,D_{\rm KL}(p\Vert \tfrac{1}{2})$, and one finds \cite{kehagias-1}
\begin{equation}
  S_{\rm bh}(M) = D_{\rm KL}^{(N)}\!\left(p\Big\Vert\frac{1}{2}\right)
               = N\,D_{\rm KL}\!\left(p\Big\Vert\frac{1}{2}\right).
\label{eq:S-as-KL}
\end{equation}
The black-hole entropy is therefore the relative entropy between the
mass-biased Bernoulli ensemble and the unbiased reference ensemble of
$N$ independent bits. It also has a large-deviation interpretation.
If $\widehat p_N$ denotes the empirical fraction of positive outcomes in
a sample of $N$ bits drawn from the unbiased distribution, then  for large
$N$, the probability of observing an empirical bias close to $p$ behaves
as
\begin{equation}
\operatorname{Prob}_{1/2}
\left(
\widehat p_N \simeq p
\right)
\sim
\exp\left[
-ND_{\rm KL}\!\left(
p\middle\Vert\frac12
\right)
\right]
=
\exp\left(-S_{\rm bh}\right). 
\end{equation}
 Thus, $S_{\rm bh}$ governs the suppression of fluctuations in which
the unbiased reference ensemble displays the mass-dependent empirical
bias $p$. Equivalently, it quantifies the statistical distinguishability
between the black-hole distribution and the uniform reference
distribution.

\section{Gravity Theory Underlying the Information-Theoretic Entropy}
\label{sec:further}

We now turn to two  questions raised by the entropy formula \eqref{sm5}. The first question concerns the infrared gravitational theory.
 Since general relativity, together with the
  Bekenstein-Hawking area law, does not reconcile black-hole
  thermodynamics with the Nernst formulation of the third law, what
  modification of the gravitational theory leads to the entropy
  \(S_{\rm bh}(M)\) given in Eq.~\eqref{sm5}?
The second question is about charged and/or rotating black holes.   Can the entropy formula \eqref{sm5} be
  generalized to describe rotating and/or charged black holes, and how should
  it be written in those cases?

\subsection{Gravity modification}

The entropy described in the previous sections suggests a corresponding
modification of the gravitational dynamics. In Einstein gravity, the
Schwarzschild solution yields the Bekenstein--Hawking entropy and the
standard Hawking temperature, whose zero-temperature behavior does not
satisfy the Nernst requirement adopted here. We therefore seek an
infrared-deformed gravitational theory whose surface-gravity temperature
reproduces the temperature implied by the information-theoretic entropy and
whose Wald entropy agrees with the same entropy.

We restrict attention to static, spherically symmetric spacetimes and adopt
the  metric ansatz \cite{Visser:1992qh}
\begin{equation}
  \mathrm{d}s^2
  = - e^{-2\phi(r)}
      \left(1 - \frac{b(r)}{r}\right)\mathrm{d}t^2
    + \frac{\mathrm{d}r^2}{1 - \dfrac{b(r)}{r}}
    + r^2 \mathrm{d}\Omega^2,
\label{eq:metric-ansatz}
\end{equation}
where $\phi(r)$ and $b(r)$ are two functions to be determined from the field
equations.  The entropy of such a background should be given by Eq. \eqref{sm3}, 
whose small–$M/M_0$ expansion reads
\begin{equation}
S_{\rm bh}
= 4\pi G_N M^2+\frac{2\pi G_N M^4}{3 M_0^2}+\frac{4\pi G_N M^6}{15 M_0^4}+\cdots\, .
\label{eq:T-expansion}
\end{equation}
For $M/M_0\ll1$ the first term is reproduced by the Schwarzschild result, while the
subleading terms encode infrared deviations from General Relativity controlled
by the universal mass scale $M_0$.
To produce such an entropy  relation like the one in \eqref{eq:T-expansion} from a gravitational theory, we
consider an effective action of the schematic form
\begin{equation}
  I = \frac{1}{16\pi G_N}\int \mathrm{d}^4x\,\sqrt{-g}\,
  \Bigl(
    R - c_0M_0^{-2} + c_1 M_0^{-4}\mathcal{I}_1
    + c_2 M_0^{-6}\mathcal{I}_2
    + \cdots
  \Bigr),
  \label{eq:effective-action}
\end{equation}
where $\mathcal{I}_i$ are functions of  higher curvature invariants such as
$R_{\mu\nu\rho\sigma}R^{\mu\nu\rho\sigma}$, $R_{\mu\nu}R^{\mu\nu}$, $R^2$,
and so on.  The coefficients $c_i$ are chosen to scale with appropriate
powers of the universal mass $M_0$ so that the Einstein–Hilbert term
dominates in the limit $M_0\to\infty$, while the $\mathcal{I}_i$ terms
generate corrections suppressed by powers of $1/M_0^2$.  In this sense the
theory represents an infrared deformation of General Relativity, where the extra terms  become important only when the mass $M$ approaches the universal
scale $M_0$, i.e.\ in the regime relevant for the Nernst limit $T\to0$.

It is well known that adding positive powers of curvature to the
Einstein–Hilbert action, such as  or $R_{\mu\nu\rho\sigma}R^{\mu\nu\rho\sigma}$, typically
induces corrections to black–hole thermodynamics that scale as inverse powers
of the black–hole mass \cite{Visser:1992qh,Lu:1993sq}.  In particular, the Hawking temperature receives
terms of order $1/M^3, 1/M^5$, etc., reflecting the fact that such operators
are important at large curvatures (small $M$) and negligible in the infrared.
In our case, however, the information-theoretic entropy \eqref{eq:T-expansion} contains
corrections that are \emph{proportional} to positive powers of $M$ rather
than inverse powers.  This behaviour cannot be generated by conventional
higher–curvature terms alone.  Consequently, the effective action that
underlies our entropy formula must also involve operators that are
nonanalytic in the curvature, or equivalently terms that behave as inverse
powers of curvature invariants in the regime of interest~\cite{Carroll:2004de}.  Such inverse–curvature
contributions are naturally associated with infrared modifications of gravity
and provide precisely the kind of mass–enhanced corrections to the
entropy required by the third law-consistent thermodynamics developed in
this work.

The full field equations derived from \eqref{eq:effective-action} are highly
nonlinear and, for generic choices of the invariants $\mathcal{I}_i$, are not
solvable in closed analytic form.  Instead, we look for solutions of the
static, spherically symmetric form~\eqref{eq:metric-ansatz} perturbatively in
the couplings $c_i$.  Concretely, we expand the metric functions as
\begin{align}
  b(r) = & b_0(r) + \frac{1}{M_0^2} b_1(r) + \frac{1}{M_0^4} b_2(r) + \cdots,
  \nonumber \\
  \phi(r) = & \phi_0(r) + \frac{1}{M_0^2} \phi_1(r) + \frac{1}{M_0^4} \phi_2(r)
  + \cdots,
  \label{eq:metric-expansion}
\end{align}
where $b_0(r)$ and $\phi_0(r)$ correspond to the Schwarzschild solution of
Einstein gravity, whereas $b_i$ and $\phi_i$  are of order $c_{i-1}$.  Inserting these expansions into the modified field
equations to linear order in the couplings $c_i$ determines $b_i(r)$, $\phi_i(r)$, and 
$r_H(M)$, and hence the entropy 
through the Wald  Noether-charge formula
\cite{Wald:1993nt,Iyer:1994ys},
\begin{equation}
S_W
=
-2\pi
\int_{\mathcal H} d^2x\,\sqrt{h}\,
\frac{\partial \mathcal L}
{\partial R_{\mu\nu\rho\sigma}}
\epsilon_{\mu\nu}\epsilon_{\rho\sigma}.
\label{eq:Wald}
\end{equation}
where $\mathcal H$ denotes a spatial cross section of the event horizon,
located at $r=r_H$.
Then, the Wald entropy (\ref{eq:Wald}) is identified with the information-theoretic entropy of Eq. (\ref{sm3}) 
 Hence, the
couplings $c_i$ in \eqref{eq:effective-action} are  fixed by demanding
that the perturbative expansion of $S$ matches the series
\eqref{eq:T-expansion}, which in turn reproduces perturbatively the entropy formula of
the previous section.

\subsection{A concrete infrared deformation }

For definiteness, let us now consider one explicit effective action that
reproduces the entropy  \eqref{eq:T-expansion}, namely 
\begin{equation}
  I = \frac{1}{16\pi G_N}\int \mathrm{d}^4x\,\sqrt{-g}\,
  \Bigl(
     R - c_0\,M_0^{-2} + c_1\,M_0^{-4}\,\mathcal{I}_1+\cdots
  \Bigr),
  \label{eq:IR-action-specific}
\end{equation}
where $c_0$ and $c_1$ are couplings with mass dimensions four and eight, respectively, and $\mathcal{I}_1$ is a curvature invariant built from the
Riemann tensor.  A simple representative choice with the correct dimensions, sufficient for our purposes,
is
\begin{equation}
  \mathcal{I}_1
  = \frac{R_{\mu\nu\rho\sigma}R^{\mu\nu\rho\sigma}}{R_{\mu\nu\rho\sigma}
{R^{\rho\sigma}}_{\kappa\lambda}R^{\kappa\lambda\mu\nu}}.
  \label{eq:I1-def}
\end{equation}
The constant term $c_0$
provides an effective cosmological constant that is tuned to vanish in
the $M_0\to\infty$ limit, while $c_1$ controls the leading higher curvature
term.  

We again adopt the static, spherically symmetric ansatz
\begin{equation}
  \mathrm{d}s^2
  = - e^{-2\phi(r)}
      \left(1 - \frac{b(r)}{r}\right)\mathrm{d}t^2
    + \frac{\mathrm{d}r^2}{1 - \dfrac{b(r)}{r}}
    + r^2 \mathrm{d}\Omega^2,
  \label{eq:metric-ansatz-again}
\end{equation}
and we look for solutions perturbative in the couplings $c_0$ and $c_1$.  It
is convenient to organize the metric functions as in Eq. \eqref{eq:metric-expansion}
where $b_1$, $b_2$ and $\phi_1$ are linear in $c_0$ and $c_1$, whereas
the leading terms $b_0$ and $\phi_0$ reproduce the Schwarzschild solution of
Einstein gravity, i.e., 
\begin{equation}
  b_0(r) = 2G_NM,
  \qquad
  \phi_0(r) = 0,
  \label{eq:b0-phi0}
\end{equation}
In other words, at zeroth order  one recovers the usual Schwarzschild
metric with horizon radius $r_H^{(0)}=2G_NM$ and Hawking temperature
$T_H=1/(8\pi G_N M)$.
Solving the field equations derived from
\eqref{eq:IR-action-specific}--\eqref{eq:I1-def} perturbative in the couplings $c_0$ and $c_1$, we find  the following corrections
to the Schwarzschild geometry
\begin{align}
  b_1(r) &= \frac{c_0}{6}\,r^3,
  \label{eq:b1-def}\\[2pt]
  b_2(r) &= \frac{19c_1}{8G_N M }\,r^6
           - \frac{9c_1}{8 G_N^2 M^2}\,r^7,
  \label{eq:b2-def}\\[2pt]
  \phi_1(r)&=0,\\[2pt]
  \phi_2(r) &= \frac{9c_1}{16G_N^2M^2}\,r^6.
  \label{eq:phi1-def}
\end{align}
Therefore, the solution for the unknown $b(r)$ and $\phi(r)$ are
\begin{align}
  b(r)
  &= 2G_NM
   \;+\; \frac{c_0}{6M_0^2}\,r^3
   \;+\; \frac{19c_1}{8G_NM M_0^4}\,r^6
   \;-\; \frac{9c_1}{8G_N^2M^2 M_0^4}\,r^7
   \;+\; \cdots,
  \label{eq:b-solution}\\[4pt]
  \phi(r)
  &= \frac{9c_1}{16G_N^2M^2 M_0^4}\,r^6
   \;+\; \cdots,
  \label{eq:phi-solution}
\end{align}
from where we find that the horizon is at 
\begin{equation}
    r_H=2G_N M\left[1+\frac{2}{3} G_N^2 M^2\left(\frac{c_0}{M_0^2}+\frac{6c_1}{M_0^4}G_N^2 M^2\right)+\cdots\right].
    \label{eq:rh}
\end{equation}

Evaluating now the Wald entropy \eqref{eq:Wald}, we find that 
\begin{align}
&S_W
=
\frac{\pi r_H^2}{G_N}+S_1
\end{align}
where 
\begin{align}
    S_1=-
\frac{c_1}{8G_NM_0^4}
\int_{\mathcal H}d^2x\,\sqrt{h}\,
\left(
\frac{2R^{\mu\nu\rho\sigma}}{R_{\alpha\beta\gamma\delta}
{R^{\gamma\delta}}_{\lambda\kappa}
R^{\lambda\kappa\alpha\beta}}
-
\frac{3(R_{\alpha\beta\gamma\delta}
R^{\alpha\beta\gamma\delta})}{(R_{\alpha\beta\gamma\delta}
{R^{\gamma\delta}}_{\lambda\kappa}
R^{\lambda\kappa\alpha\beta})^2}
{R^{\rho\sigma}}_{\epsilon\zeta}
R^{\epsilon\zeta\mu\nu}
\right)
\epsilon_{\mu\nu}\epsilon_{\rho\sigma}, 
\label{eq:Wald-explicit}
\end{align}
which, using Eqs (\ref{eq:b-solution}) and (\ref{eq:phi-solution})  
turns out to be 
\begin{equation}
S_W
=
4\pi G_N M^2
+
\frac{16\pi c_0 G_N^3 M^4}{3M_0^2}
-
\frac{32\pi c_1 G_N^5 M^6}{3M_0^4}
+
\mathcal O(c_i^2).
\label{eq:Wald-expansion}
\end{equation}
Hence, matching Eqs (\ref{eq:T-expansion}) and (\ref{eq:Wald-expansion}), we find that the two expressions coincide, i.e.,
\begin{equation}
    S_{\rm bh}=S_W
\end{equation}
for the following choice of the couplings
\begin{eqnarray}
    c_0=\frac{1}{8G_N^2}, 
    \qquad\mbox{and}\qquad c_1=-\frac{1}{40 G_N^4 },
    \label{eq:c}
\end{eqnarray}
  and  the effective
gravity theory \eqref{eq:IR-action-specific} turns out to be 
\begin{equation}
  I = \frac{1}{16\pi G_N}\int \mathrm{d}^4x\,\sqrt{-g}\,
  \bigg(
     R - \frac{1}{8 G_N^2M_0^{2}} -\frac{1}{40 G_N^4M_0^{4}}\,\frac{R_{\mu\nu\rho\sigma}R^{\mu\nu\rho\sigma}}{R_{\mu\nu\rho\sigma}
{R^{\rho\sigma}}_{\kappa\lambda}R^{\kappa\lambda\mu\nu}}+\cdots
  \bigg).
  \label{eq:IR-action-specific1}
\end{equation}
Thus, to linear order in the deformation couplings, the Wald
Noether-charge entropy of the effective theory agrees with the
information-theoretic entropy. 
In addition, 
the Hawking temperature obtained from the surface gravity is \cite{Visser:1992qh,Lu:1993sq}
\begin{equation}
T
=
\left.
\frac{e^{-\phi(r)}}{4\pi r}
\left[1-b'(r)\right]
\right|_{r=r_H}.
\end{equation}
Using Eqs.~\eqref{eq:b-solution}, \eqref{eq:phi-solution}, and
\eqref{eq:rh}, and retaining only terms linear in $c_0$ and $c_1$, gives
\begin{equation}
T
=
\frac{1}{8\pi G_NM}
\left[
1
-\frac{8c_0G_N^2M^2}{3M_0^2}
+\frac{8c_1G_N^4M^4}{M_0^4}
\right]
+
\mathcal O(c_i^2),
\label{eq:T-surface-gravity}
\end{equation}
which coincides with $T=(dS_W/dM)^{-1}$ where $S_W$ is given by eq. \eqref{eq:Wald-expansion}.
The above agreements provide a consistency check between the
thermodynamic, geometrical, and information-theoretic descriptions. The
result holds within the perturbative truncation considered here, since
terms nonlinear in the couplings and additional higher-order operators
have not been included.

We emphasize that the solution
\eqref{eq:b-solution}--\eqref{eq:phi-solution} should be understood as a
perturbative deformation of Schwarzschild, valid in the regime
$M\ll M_0$ and at distances $r$ not too close to the curvature scale set by
$M_0$.  It provides a concrete example of how infrared modifications of
general relativity can realize the thermodynamic properties required by the
Nernst theorem while smoothly interpolating back to standard General Relativity in the
appropriate limit.

One may ask to what extent the action \eqref{eq:effective-action}, or \eqref{eq:IR-action-specific1}, is unique.
From the viewpoint of effective field theory, the answer is that it is not unique in a
strict sense.  Many different combinations of higher–curvature invariants,
possibly related by field redefinitions or differing by terms that vanish on
the spherically symmetric background, can lead to the \emph{same} temperature
and entropy for static, neutral black holes, at least up to the order kept in
the $1/M_0$ expansion.  In particular, in our explicit construction we have
concentrated on invariants built from the Riemann tensor, such as
$R_{\mu\nu\rho\sigma}R^{\mu\nu\rho\sigma}$, and neglected terms involving the
Ricci scalar $R$ or the Ricci tensor $R_{\mu\nu}$.  For the Schwarzschild
vacuum solution these quantities vanish at leading order, so operators built
from $R$ or $R_{\mu\nu}$ contribute only at higher orders in the perturbation
around General Relativity and can be consistently dropped within the accuracy of our
analysis.  Our construction should therefore be regarded as an existence
proof. It shows that there do exist infrared deformations of General Relativity whose
black–hole solutions reproduce the desired thermodynamic behaviour and the
third–law–consistent entropy.  Additional physical input (such as
requirements from cosmology, scattering amplitudes, or embedding in a UV
complete theory) would be needed to single out a particular set of curvature
invariants as the ``correct'' one.  Within the level of description adopted
here, the action is best viewed as a minimal representative of a class of
theories that share the same black–hole thermodynamics.

\section{Extension to Charged and Rotating Black Holes}
\label{sec:extension}

Having established an entropy formula consistent with the third law of thermodynamics for neutral, non-rotating Schwarzschild black holes, we now turn our attention to the more general case of stationary black holes carrying angular momentum and/or electric charge.
In this setting the ADM mass $M$ is no longer the most natural variable to
use in the entropy, since $M$ includes contributions from rotational and
electromagnetic energy that can, at least in principle, 
reversibly be extracted. 
This motivates reformulating our entropy in terms of the \emph{irreducible
mass} $M_{\rm irr}$, which 
remains
invariant under reversible extraction processes ~\cite{Christodoulou:1970wf,Christodoulou:1971pcn}.  The central proposal of
this section is therefore simple. The Schwarzschild entropy formula derived
above is generalized to Kerr and 
Kerr-Newman black holes by the replacement
$M$ by $M_{\rm irr}$.  We will first recall below the definition and basic
properties of $M_{\rm irr}$ and then present the resulting entropy and
temperature formulas, discussing their limiting behavior and the
implementation of the third law in the rotating/charged sector.

\subsection{The irreducible mass}

The irreducible mass $M_{\rm irr}$ is the natural mass scale associated with
the horizon area and, consequently, with the part of the black hole that
cannot be lowered by any classical extraction process.  It is defined by
\begin{equation}
  M_{\rm irr} \;\equiv\; \sqrt{\frac{A}{16\pi G_N^2}},
\end{equation}
so that it depends only on the geometry of the horizon.  For Kerr-Newman
black holes, specified by the ADM mass $M$, angular momentum $J$, and charge
$Q$, one may express $M_{\rm irr}$ explicitly in terms of $(M,J,Q)$.  Writing
the specific angular momentum as $a \equiv J/M$ and denoting by $r_+$ the
outer horizon radius,
\begin{equation}
  r_+ \;=\; G_NM + \sqrt{G_N^2M^2 - G_NQ^2 - a^2},
\end{equation}
the horizon area is
\begin{equation}
  A \;=\; 4\pi\left(r_+^2 + a^2\right),
\end{equation}
and therefore
\begin{equation}
  M_{\rm irr}^2
  \;=\; \frac{A}{16\pi G_N^2}
  \;=\; \frac{r_+^2 + a^2}{4G_N^2}.
\label{eq:Mirr-KN-rplus}
\end{equation}
Eliminating $r_+$ and $a$ in favour of the conserved charges $(M,J,Q)$ yields
a more directly thermodynamic expression,
\begin{equation}
  M_{\rm irr}^2
=
\frac{1}{2}
\left[
M^2-\frac{Q^2}{2G_N}
+
\sqrt{
M^4-\frac{M^2Q^2}{G_N}-\frac{J^2}{G_N^2}
}
\right],
\label{eq:Mirr-KN-MJQ}
\end{equation}
which reduces to the familiar Kerr and Reissner-Nordstr\"om limits when
$Q=0$ or $J=0$, respectively.

The significance of $M_{\rm irr}$ is best understood through the
Christodoulou-Ruffini mass formula~\cite{Christodoulou:1971pcn}.  For a Kerr-Newman black hole one may
decompose the ADM mass into an ``irreducible'' contribution plus an
``extractable'' part associated with rotation and electromagnetic energy, arriving in the Christodoulou-Ruffini mass formula
\cite{Christodoulou:1970wf,Christodoulou:1971pcn}
\begin{equation}
  M^2
  \;=\;
  \left(M_{\rm irr} + \frac{Q^2}{4G_N\,M_{\rm irr}}\right)^{\!2}
  + \frac{J^2}{4G_N^2\,M_{\rm irr}^2}.
\label{eq:Christodoulou-Ruffini}
\end{equation}
This relation makes it clear that, at fixed $M_{\rm irr}$, adding charge or
angular momentum increases the ADM mass by storing energy in the external
electromagnetic field and in rotation.  Conversely, by allowing interactions
with external agents one may \emph{decrease} $M$ by extracting rotational
energy (Penrose processes, superradiance) and by discharging the black hole,
while keeping the horizon area unchanged in an ideal reversible limit.
This is precisely why $M_{\rm irr}$ is called ``irreducible'' since classical
processes can reduce the ADM mass down to a minimum determined by $M_{\rm irr}$
but cannot reduce $M_{\rm irr}$ itself.  In fact, Hawking's area theorem
implies that in any classical process obeying suitable energy conditions the
horizon area cannot decrease, and therefore
\begin{equation}
  \Delta A \ge 0
  \qquad \Longrightarrow \qquad
  \Delta M_{\rm irr} \ge 0.
\end{equation}
Thus $M_{\rm irr}$ behaves as a monotone measure of the horizon's
non-decreasing ``core'' size.  This property is the precise analogue of the
statement that entropy should not decrease in ordinary thermodynamics, and it
is the reason why the Bekenstein-Hawking entropy depends only on $M_{\rm irr}$:
\begin{equation}
  S_{\rm BH} = \frac{A}{4G_N} = 4\pi G_N\,M_{\rm irr}^2.
\end{equation}

These considerations provide the physical justification for replacing $M$ by
$M_{\rm irr}$ in our third-law-consistent entropy formula for charged and/or
rotating black holes.  The ADM mass $M$ includes not only the irreducible
horizon contribution but also energy stored in degrees of freedom that, at
least in principle, can be extracted reversibly and need not be associated
with an increase in horizon entropy.  By contrast, $M_{\rm irr}$ is tied
directly to the horizon area, is invariant under reversible extraction
processes, and is the unique mass parameter that captures the portion of the
black hole that cannot be reduced without decreasing the horizon area.  If
our entropy is to quantify the amount of information irretrievably hidden
behind the horizon (or, in the microscopic picture, the number of hidden
microstates associated with the horizon degrees of freedom), then it is
natural that it should depend on $M_{\rm irr}$ alone.  In this way, the
generalization
\begin{equation}
  S_{\rm bh}(M)\;\longrightarrow\; S_{\rm bh}(M_{\rm irr})
\end{equation}
extends our construction to Kerr and Kerr-Newman black holes while preserving
its central conceptual feature: the entropy is a function of the
area-carrying, non-extractable component of the black hole, rather than of
the total ADM energy.


\subsection{Entropy and Temperature}

Based on the above, the proposed generalization of our third-law-consistent entropy is simply
obtained by the replacement $M$ by $M_{\rm irr}$ in our Schwarzschild formula so that 
\begin{align}
     S_{\bh}= &  \frac{N}{2}\bigg[\left(1-\frac{M_{\rm irr}}{M_0}\right)\ln \left(1-\frac{M_{\rm irr}}{M_0}\right)+\left(1+\frac{M_{\rm irr}}{M_0}\right)\ln\left(1+\frac{M_{\rm irr}}{M_0}\right)\bigg].
     \label{eq:Sbh-KN-Mirr}
 \end{align}
As before, $M_0$ is the universal mass scale controlling the infrared
deviation from General Relativity and $N=M_0^2/M_P^2$ is the total number of microscopic bits in the universe.
Moreover, it is useful to note the derivative of $S_{\bh}$ with respect to $M_{\rm irr}$, which will
enter the temperature below, is given by
\begin{equation}
\frac{\partial S_{\rm bh}}{\partial M_{\rm irr}}
=
\frac{N}{2M_0}\,
\ln\!\left(\frac{M_0+M_{\rm irr}}{M_0-M_{\rm irr}}\right).
\label{eq:dS-dMirr}
\end{equation}

To calculate the associated temperature now, we note that for Kerr-Newman black holes the first law reads
\begin{equation}
dM = T\,dS + \Omega\,dJ + \Phi\,dQ.
\end{equation}
Therefore, at fixed $J$ and $Q$ the Hawking temperature is
\begin{equation}
T(M,J,Q)
=
\left(\frac{\partial M}{\partial S}\right)_{J,Q}
=
\frac{\left(\frac{\partial M}{\partial M_{\rm irr}}\right)_{J,Q}}
     {\left(\frac{\partial S_{\rm bh}}{\partial M_{\rm irr}}\right)}.
\label{eq:T-general}
\end{equation}
Using \eqref{eq:Christodoulou-Ruffini}, we find that the derivative $(\partial M/\partial M_{\rm irr})_{J,Q}$ is given by 
\begin{equation}
\left(\frac{\partial M}{\partial M_{\rm irr}}\right)_{J,Q}
=
\frac{1}{M}\left[
M_{\rm irr}
-\frac{1}{16G_N^2}\,\frac{Q^4+4J^2}{M_{\rm irr}^3}
\right].
\label{eq:dM-dMirr}
\end{equation}
Combining \eqref{eq:dS-dMirr} and \eqref{eq:dM-dMirr} in \eqref{eq:T-general},
we obtain an explicit closed expression for the temperature 
\begin{equation}
T(M,J,Q)
=
\frac{2M_0}{N}\;
\frac{
\displaystyle
\left[
M_{\rm irr}
-\frac{Q^4+4J^2}{16G_N^2\,M_{\rm irr}^3}
\right]
}{
\displaystyle
M\;
\ln\!\left(\frac{M_0+M_{\rm irr}}{M_0-M_{\rm irr}}\right)
},
\label{eq:T-KN-ourmodel}
\end{equation}
where it is  understood that $M_{\rm irr}=M_{\rm irr}(M,J,Q)$  by \eqref{eq:Mirr-KN-MJQ}.

\paragraph{Low-mass (GR) limit.}
For $M_{\rm irr}\ll M_0$ we have
\begin{equation}
\ln\!\left(\frac{M_0+M_{\rm irr}}{M_0-M_{\rm irr}}\right)
= \frac{2M_{\rm irr}}{M_0} + \mathcal{O}\!\left(\frac{M_{\rm irr}^3}{M_0^3}\right),
\end{equation}
so that from Eq. (\ref{eq:dS-dMirr}) we find 
\begin{equation}
\frac{\partial S_{\rm bh}}{\partial M_{\rm irr}}
=
\frac{N}{M_0^2}\,M_{\rm irr}
+\mathcal{O}\!\left(\frac{M_{\rm irr}^3}{M_0^4}\right). \label{S-p}
\end{equation}
Therefore, with $N=M_0^2/M_P^2=8\pi G_N M_0^2$ Eq. (\ref{S-p}) yields
\begin{equation}
S_{\rm bh}
=
4\pi G_NM_{\rm irr}^2
\left[
1
+
\frac{M_{\rm irr}^2}{6M_0^2}
+
\mathcal{O}
\left(
\frac{M_{\rm irr}^4}{M_0^4}
\right)
\right].
\end{equation}
which is at leading order, the familiar Hawking entropy of a charged, rotating black hole.
Similarly, the temperature turns out to be
\begin{equation}
T
=
\frac{1}{8\pi G_NM}
\left(
1-
\frac{Q^4+4J^2}
{16G_N^2M_{\rm irr}^4}
\right)
\left[
1-
\frac{M_{\rm irr}^2}{3M_0^2}
+
\mathcal{O}
\left(
\frac{M_{\rm irr}^4}{M_0^4}
\right)
\right].
\end{equation}

In the present framework, one zero-temperature endpoint occurs for
$$
M_{\rm irr}\to M_0.
$$
At this endpoint the logarithm in \eqref{eq:T-KN-ourmodel} diverges and,
provided the mechanical prefactor remains finite,  we have 
\begin{equation}
T(M,J,Q)\xrightarrow[M_{\rm irr}\to M_0]{}0.
\end{equation}
At the same time, the entropy approaches
\begin{equation}
\lim_{M_{\rm irr}\to M_0}S_{\rm bh}(M,J,Q)
=N\ln 2\equiv S_0,
\end{equation}
independently of $(J,Q)$. Thus, along this statistical zero-temperature
endpoint, the entropy approaches the universal constant required by the
Nernst formulation of the third law. The Hawking temperature can also
vanish for a different reason, namely at geometrical extremality. This
second zero-temperature limit is discussed in the following subsection.


\subsection{The extremal limit and the third law} \label{sec:extremal-third-law}

The temperature \eqref{eq:T-KN-ourmodel} has two distinct zeros. Since the entropy depends on $M$, $J$, and $Q$ only through the irreducible mass, the temperature can be written as \begin{equation} T = \frac{\Delta}{M}\,T_{\rm ens}, 
\label{eq:T-factorized}
\end{equation}
where
\begin{equation}
    T_{\rm ens}(M_{\rm irr}) = \left( \frac{\partial S_{\rm bh}}{\partial M_{\rm irr}} \right)^{-1} = \frac{2M_0}{N} \left[ \ln \left( \frac{M_0+M_{\rm irr}}{M_0-M_{\rm irr}} \right) \right]^{-1},  \end{equation} and  \begin{equation} \Delta = M_{\rm irr} - \frac{Q^4+4J^2} {16G_N^2M_{\rm irr}^3}. \end{equation} The factor $\Delta/M$ in eq. \eqref{eq:T-factorized} is mechanical and vanishes at extremality, where the inner and outer horizons coincide. The quantity $T_{\rm ens}$ is instead the statistical temperature of the microscopic Bernoulli ensemble. It is finite for $0<M_{\rm irr}<M_0$ and vanishes only in the limit $M_{\rm irr}\to M_0$.
    Therefore, the vanishing of $T$ at extremality is caused by the mechanical
factor $\Delta/M$, which reflects the degeneracy of the horizon geometry.
 The microscopic distribution still has nonzero
Shannon entropy and nonvanishing fluctuations in $M_{\rm irr}$ whenever
$M_{\rm irr}^{\rm ext}<M_0$. In this sense, the extremal state is
geometrically cold but not statistically frozen.
    This distinction is also visible in the fluctuations of the irreducible mass. 
    More precisely, defining the Bernoulli estimator
$\widehat M_{\rm irr}=M_0(2\widehat p-1)$, where $\widehat p$ is the sample
mean of the $N$ binary variables, one has
$\langle\widehat M_{\rm irr}\rangle=M_{\rm irr}$ and
    \begin{equation} \left\langle \left(\delta M_{\rm irr}\right)^2 \right\rangle = M_P^2 \left( 1-\frac{M_{\rm irr}^2}{M_0^2} \right), \label{eq:Mirr-fluct} \end{equation} 
    where $\delta M_{\rm irr}=\widehat M_{\rm irr}-M_{\rm irr}$.
    These fluctuations remain finite on the extremal branch and vanish only when $M_{\rm irr}\to M_0$. Extremality therefore suppresses variations of the ADM mass through the vanishing of the mechanical factor $\Delta/M$, but it does not freeze the microscopic horizon ensemble. The interpretation of the entropy is essential here. In the present model, \begin{equation} S_{\rm bh} = D_{\rm KL}^{(N)} \left( p({M_{\rm irr}}) \middle\| \frac{1}{2} \right),\end{equation} and thus $S_{\rm bh}$ is a relative entropy. It measures the information that distinguishes the biased black hole ensemble from the unbiased reference ensemble. It is not the logarithm of a number of internal states. The extremal limit must therefore be distinguished from the
zero-temperature limit of the Bernoulli ensemble. At geometric
extremality, the Hawking temperature vanishes because
$\Delta/M\to0$, while the ensemble temperature remains finite whenever
$M_{\rm irr}^{\rm ext}<M_0$. Geometric extremality is therefore not the
limit $T_{\rm ens}\to0$ in which the Nernst condition is implemented in
the present model. The finite value
$S_{\rm bh}(M_{\rm irr}^{\rm ext})$ should consequently not, by itself,
be interpreted as the logarithm of a degeneracy of exact ground states.
The fate of the apparent parameter-dependent extremal entropy requires
the quantum near-horizon analysis discussed below. This
interpretation is consistent with quantum analyses of near-extremal black
holes, in which fluctuations of the near-horizon region modify the
semiclassical extremal spectrum. For nonsupersymmetric black holes, the
low-energy states form a continuous spectrum whose density vanishes at the
extremal threshold, with no isolated contribution proportional to
$\delta(E-E_{\rm ext})$
\cite{IliesiuTuriaci2020,
RakicRangamaniTuriaci2023,KapecShetaStromingerToldo2023}.
The semiclassical factor $\exp(S_{\rm ext})$ therefore normalizes a
continuous spectral density rather than necessarily counting degenerate
exact ground states. A direct application of this result to the present
infrared-deformed theory would require deriving its near-horizon quantum
effective theory.

   Supersymmetric extremal black holes require a separate interpretation.
The Strominger--Vafa calculation counts BPS states, and more general
microscopic treatments are often formulated in terms of protected
supersymmetric indices whose asymptotic growth can reproduce the
black-hole entropy
\cite{Strominger1996,DabholkarGomesMurthySen2011}.
These protected quantities are conceptually distinct from the KL entropy
considered here. They characterize a fixed BPS sector of the Hilbert
space, whereas $S_{\rm bh}$ characterizes a thermodynamic family of
Bernoulli distributions, and therefore, the Nernst interpretation obtained from the
Bernoulli ensemble  does not apply directly to protected BPS
states. Indeed, near-BPS black holes may retain a finite extremal
degeneracy separated from the excited spectrum by a mass gap
\cite{HeydemanIliesiuTuriaciZhao2021}.

This distinction is compatible with recent results on the unattainability
formulation of the third law, according to which supersymmetric
extremality cannot be reached from a nonextremal configuration in finite
time under suitable assumptions
\cite{Reall1,Reall2}. This dynamical statement is distinct from the
Nernst condition studied here.

\section{The universal scale \texorpdfstring{$M_0$}{M0}: From contingent bound to the Cosmological Constant}
\label{sec:CC}

Up to this point, the scale $M_0$ has been introduced thermodynamically as the maximal mass at which the temperature vanishes, and interpreted heuristically as the mass that would result from the collapse of the entire observable Universe. This interpretation, however, is problematic since the mass of the Universe is a contingent, epoch-dependent quantity, rather than a constant of nature. Consequently, one may worry that the entropy of a stellar black hole would thereby acquire a spurious dependence on cosmology. In this section, we demonstrate how the effective action derived in Section~3 resolves this issue. Specifically, the matching that fixed the couplings $(c_0,c_1)$ promotes $M_0$ to a fundamental parameter of the gravitational action, rigidly linked to the cosmological constant. 

The constant term in the effective action \eqref{eq:IR-action-specific1} was determined by matching the derived perturbative entropy \eqref{eq:Wald-expansion} to the  expansion \eqref{eq:T-expansion}, thereby fixing the coefficient $c_0=1/(8G_N^2)$ in Eq.~\eqref{eq:c}. Writing the action \eqref{eq:IR-action-specific1} in the standard form
$$S=\frac{1}{16\pi G_N}\int d^4x\sqrt{-g}\,(R-2\Lambda_{\rm eff}+\cdots),$$
we see that the theory contains the effective cosmological constant 
\begin{equation}
\Lambda_{\rm eff}=\frac{c_0}{2M_0^2}=\frac{1}{16\,G_N^{2}M_0^{2}} =\frac{4\pi^{2}M_P^{2}}{N}\,.
\label{eq:LambdaN}
\end{equation}
It is easy also to see that indeed, the constant term in the effective action generates a de Sitter-like
contribution to the perturbative black-hole geometry. Using
\begin{equation}
b(r)
=
2G_NM+\frac{c_0}{6M_0^2}r^3+\cdots,
\end{equation}
the metric function becomes
\begin{equation}
1-\frac{b(r)}{r}
=
1-\frac{2G_NM}{r}
-\frac{c_0}{6M_0^2}r^2+\cdots, 
\end{equation}
which is of the Schwarzschild--de Sitter form
\begin{equation}
1-\frac{2G_NM}{r}-\frac{\Lambda_{\rm eff}}{3}r^2
\end{equation}
with cosmological constant given in Eq. \eqref{eq:LambdaN}. 

Thus, within the perturbative solution and to linear order in the
deformation couplings, the leading cosmological term is controlled by
$c_0$.
Two remarks are in order here. First, since $\Lambda_{\rm eff}>0$, the
leading cosmological term has the de Sitter sign and is compatible with an
accelerating universe. Second, the smallness of the cosmological constant
is mapped onto the largeness of the parameter $N$ of the microscopic
ensemble,
\begin{equation}
\Lambda_{\rm eff}\sim\frac{M_P^2}{N}.
\label{eq:DG}
\end{equation}
This relation does not by itself solve the cosmological constant problem, but rephrase it  in a different way. It should be mentioned here that the relation \eqref{eq:DG} appeared in \cite{Dvali:2013eja}, where though $N$ represents graviton occupation number.

Hence, identifying $\Lambda_{\rm eff}$ with the observed value $\Lambda_{\rm obs}\simeq 3\Omega_\Lambda H_0^{2}\simeq 7.2\times10^{-121}\,M_P^{2}$ \cite{Planck:2018vyg} determines the two parameters of the framework with no further freedom
\begin{equation}
N=\frac{4\pi^{2}M_P^{2}}{\Lambda_{\rm obs}}\simeq 5.5\times10^{121}, 
\end{equation}
and \begin{equation}M_0=\frac{2\pi M_P^{2}}{\sqrt{\Lambda_{\rm obs}}} \simeq 7.4\times10^{60}\,M_P. 
\label{eq:NM0numbers}
\end{equation}
The physical nature of $M_0$ is therefore a constant of nature, on equal footing with $\Lambda_{\rm obs}$ itself, meaning that the entropy of a stellar black hole depends on cosmology only through utterly negligible $1/N\sim10^{-122}$ corrections.

The vacuum solution of the effective theory has a cosmological constant $\Lambda_{\rm eff}$, which is connected to de Sitter space.  The corresponding Gibbons--Hawking area quantity is \cite{GibbonsHawking}  
\begin{equation} 
S_{\rm dS}^{\rm GH} = \frac{24\pi^2 M_P^2}{\Lambda_{\rm eff}} = 6N. 
\label{eq:SdS} 
\end{equation} 
Notice that here $S_{\rm dS}^{\rm GH}$ is understood as the leading semiclassical quantity associated with the de Sitter horizon. Equation \eqref{eq:SdS} does not imply that $N$ is the microscopic entropy of de Sitter space, or that $\exp(S_{\rm dS}^{\rm GH})$ necessarily counts fundamental de Sitter states. It shows only that the same large parameter $N$ that fixes the size of the microscopic Bernoulli ensemble also fixes the semiclassical de Sitter horizon scale. At the zero-temperature endpoint, 
\begin{equation} 
S_0 = N\ln 2 = \frac{\ln 2}{6}\,S_{\rm dS}^{\rm GH}. 
\label{eq:S0SdS} 
\end{equation} 
Thus, the endpoint value of the black-hole relative entropy is proportional to the semiclassical de Sitter horizon quantity, without identifying the two as the same type of entropy.

The result of this section indicates that what originally appeared to be an arbitrary choice for the scale $M_0$ in \cite{kehagias-1} may be reinterpreted as a structural property, namely a connection between the third-law  of black hole thermodynamics and the dark energy scale.

\section{Conclusions}
\label{sec:conclusions2}

In this paper we developed further the information theoretic approach to
black hole entropy introduced in \cite{kehagias-1}, where   a modified entropy has been derived by imposing the Nernst formulation of the third
law together with the known limited behavior of the Bekenstein Hawking
result. The present analysis addressed the gravitational origin of this
entropy and its extension to charged and rotating black holes.
In particular, we showed that the modified thermodynamics can be reproduced by a class of
infrared deformations of General Relativity. These theories contain higher
curvature and inverse curvature operators controlled by the universal mass
scale $M_0$, and  the model we considered based on the invariant $\mathcal{I}_1$ provides
one explicit realization. Its coefficients were fixed by matching the
perturbative black hole temperature to the temperature implied by the
modified entropy. This construction establishes that thermodynamics
compatible with the Nernst requirement can arise from a gravitational
action, but nevertheless, it does not select a unique theory, since additional physical input
is needed to determine the full infrared completion.

The same construction generates a positive effective cosmological constant, where 
the  smallness of the latter is related to the large value of the microscopic parameter
$N$. The corresponding cosmological and black hole scales are therefore
controlled by the same infrared parameter. This relation should be regarded
as a connection between scales rather than as a microscopic interpretation
of the de Sitter horizon entropy, which has not been considered here and remains an open issue in the present framework. 

For Kerr Newman black holes, the natural thermodynamic variable is the
irreducible mass, which  is directly determined by the horizon area and remains
unchanged under reversible extraction of rotational and electromagnetic
energy. Replacing the ADM mass by the irreducible mass reproduces the
standard Kerr Newman expressions in the regime where corrections controlled
by \(M_0\) are negligible. It also gives a universal maximal entropy that is
independent of the angular momentum and charge, in accordance with Nernst third law.

The vanishing of the Hawking temperature at extremality has a distinct
geometrical origin. It results from the degeneracy of the horizon and from
the vanishing Jacobian that relates variations of the ADM mass to variations
of the irreducible mass at fixed angular momentum and charge. By contrast,
the Bernoulli ensemble retains nonzero statistical fluctuations whenever
$M_{\rm irr}<M_0$, and therefore extremality represents a geometrical
zero-temperature limit rather than the frozen microscopic endpoint selected
by the Nernst condition.

The interpretation of the entropy is central to this distinction, and in the
present framework, black hole entropy is a Kullback-Leibler divergence
between a mass biased ensemble and an unbiased reference ensemble. It
measures relative information and does not count an absolute number of
internal states. Therefore, its limiting value should not be interpreted as
the logarithm of a ground state degeneracy, but instead, it is the maximum
entropy deficit of the Bernoulli sector relative to the uniform reference.

Together with \cite{kehagias-1}, these results provide a consistent picture
in which the Bekenstein Hawking area law appears as the leading approximation
to a relative entropy, the third law introduces a universal infrared scale,
and the gravitational action is modified accordingly. The construction
remains an effective description, where a complete account will require a
microscopic dynamics for the Bernoulli variables, a better understanding of
the induced cosmological sector, and an embedding into a more fundamental
theory of quantum gravity.

\vskip.2in
\noindent\\
{\bf\large Appendix}
\vskip.04in
\appendix

\section{Compatibility with the Nariai bound}

The positive cosmological constant generated by the effective theory also
introduces a geometrical upper bound on the Schwarzschild--de Sitter mass
parameter. For the metric function
\begin{equation}
f(r)
=
1-\frac{2G_N M}{r}-\frac{\Lambda r^2}{3},
\end{equation}
the black-hole and cosmological horizons coincide at the Nariai limit, which 
correspond to the maximal mass 
\begin{equation}
M_{\rm N}
=
\frac{1}{3G_N\sqrt{\Lambda}}.
\end{equation}
Using the induced cosmological constant in Eq.~\eqref{eq:LambdaN}, one finds
\begin{equation}
M_{\rm N}
=
\frac{4}{3}M_0.
\label{eq:Nariai}
\end{equation}

The microscopic bound $M\leq M_0$ is therefore compatible with the
Schwarzschild--de Sitter geometry, since every mass allowed by the
Bernoulli model lies below the Nariai limit. The microscopic bound is in
fact slightly stronger than the geometrical one. However, the endpoint
$M=M_0$ does not coincide with the Nariai solution. Equation
\eqref{eq:Nariai} should therefore be interpreted as a compatibility
between the two upper-mass scales, not as a geometrical derivation of the
bound $M\leq M_0$.

\bibliographystyle{JHEP}
\bibliography{biblio}

\end{document}